\documentclass[prb,twocolumn,superscriptaddress,showpacs,amsmath,amssymb]{revtex4-1}

\usepackage{graphicx}
\usepackage{latexsym}
\usepackage{amsmath}
\usepackage{amssymb}
\usepackage{amsfonts}
\usepackage{color}
\usepackage{bm}
\usepackage{verbatim}
\usepackage[]{color}
\definecolor{red}{rgb}{1,0,0}
\usepackage{framed}
\definecolor{shadecolor}{RGB}{222,222,221}
\definecolor{MS-color}{RGB}{128,0,128}

\begin{document}

\title{Electrical control of magnetization in S/F/S junctions on a 3D topological insulator.}

 \date{\today}
 
\author{M. Nashaat}
\affiliation{BLTP, Joint Institute for Nuclear Research, Dubna, Moscow Region, 141980, Russia}
\affiliation{Department of Physics, Cairo University, Cairo, 12613, Egypt}

\author{I. V. Bobkova}
\affiliation{Institute of Solid State Physics, Chernogolovka, Moscow
  reg., 142432 Russia}
\affiliation{Moscow Institute of Physics and Technology, Dolgoprudny, 141700 Russia}

\author{A. M. Bobkov}
\affiliation{Institute of Solid State Physics, Chernogolovka, Moscow reg., 142432 Russia}

\author{Yu. M. Shukrinov}
\affiliation{BLTP, Joint Institute for Nuclear Research, Dubna, Moscow Region, 141980, Russia}
\affiliation{Dubna State University, Dubna,  141980, Russia}

\author{I. R. Rahmonov}
\affiliation{BLTP, Joint Institute for Nuclear Research, Dubna, Moscow Region, 141980, Russia}
\affiliation{Umarov Physical Technical Institute, TAS, Dushanbe, 734063, Tajikistan}

\author{K. Sengupta}
\affiliation{School of Physical Sciences, Indian Association for the Cultivation of Science, Jadavpur, Kolkata-700032, India}

\begin{abstract}
Strong dependence of the Josephson energy on the magnetization orientation in Josephson junctions with ferromagnetic interlayers and spin-orbit coupling opens a way to control magnetization by Josephson current or Josephson phase. Here we investigate the perspectives of magnetization control in superconductor/ferromagnet/superconductor (S/F/S) Josephson junctions on the surface of a 3D topological insulator hosting Dirac quasiparticles. Due to the spin-momentum locking of these Dirac quasiparticles a strong dependence of the Josephson current-phase relation on the magnetization orientation is realized. It is demonstrated that this can lead to splitting of the ferromagnet's easy-axis in the voltage driven regime.  We show that such a splitting can lead to stabilization of an unconventional four-fold degenerate ferromagnetic state.
\end{abstract}

 \pacs{} \maketitle
 
\section{Introduction}

By now it is well-known that current-phase relation (CPR) in Josephson junctions with multi-layered ferromagnetic interlayers is strongly sensitive to the mutual orientation of the magnetizations in the layers \cite{Waintal2002,Barash2004,Braude2007,   
Grein2009,  Liu2010, 
Kulagina2014,Moor2015_1,Moor2015_2,Mironov2015,
Silaev2017,Bobkova2017,Rabinovich2018}. CPRs of Josephson junctions with ferromagnetic interlayers in the presence of spin-orbit coupling also depends on the magnetization orientation. This occurs primarily via the appearance of the magnetization-dependent anomalous phase shift \cite{Krive2004,Asano2007,Reynoso2008,Buzdin2008,Tanaka2009,Zazunov2009,Malshukov2010,Alidoust2013,Brunetti2013,Yokoyama2014,Bergeret2015,Campagnano2015,Konschelle2015,Kuzmanovski2016}. This coupling  between the Josephson and magnetic subsystems leads to the supercurrent-induced magnetization dynamics \cite{Konschelle2009,Kulagina2014,Waintal2002,Braude2008,Linder2011,Cai2010,Chudnovsky2016,Bobkova2018}. In particular, the reversal of the magnetic moment by the supercurrent pulse \cite{Shukrinov2017} was predicted. A unique possibility of controlling the magnetization dynamics via external bias current and series of specific magnetization trajectories has been reported \cite{Shukrinov2019}. In Refs.~\onlinecite{Konschelle2009,Shukrinov2018} it was also reported that in the presence of spin-orbit coupling the supercurrent can cause reorientation of the magnetization easy-axis. Assuming the initial position of the easy axis along $z$-direction these works demonstrate that under the applied supercurrent stable position of the magnetization becomes between $z$−
and $y$-axes depending on parameters of the system. 

Here we investigate prospects of superconductor/ferromagnet/superconductor (S/F/S) Josephson junctions  constructed atop a three dimensional topological insulator (3D TI) surface, which hosts Dirac quasiparticles, in the field of supercurrent-induced magnetization control. Our motivation is that these  Dirac quasiparticles on the surface of the 3D TI exhibit full spin-momentum locking: an electron spin always makes a right angle with its momentum. This gives rise to a very pronounced dependence of the CPR on the magnetization direction  \cite{Tanaka2009,Linder2010,Zyuzin2016}. In particular, the anomalous ground state phase shift proportional to the in-plane magnetization component perpendicular to the supercurrent direction was reported. 

The second reason to study magnetization dynamics in such a system is that at present there is a great progress in experimental realization of F/TI hybrid structures. In particular, to introduce the ferromagnetic order into the TI, random doping of transition metal elements, e.g., Cr or V, has been employed \cite{Chang2013,Kou2013,Kou2013_2,Chang2015}. The second option, which has been successfully realized experimentally, is a coupling of the nonmagnetic TI to a high $T_c$
magnetic insulator to induce strong exchange interaction in the surface states via the proximity effect\cite{Jiang2015,Wei2013,Swartz2012,Jiang2015_2,Jiang2016}.

Here we demonstrate that the anomalous phase shift causes the magnetization dynamics analogously to the case of a spin-orbit coupled system. However, in contrast to the spin-orbit coupled systems, where the magnetization dynamics was studied before, for the system under consideration the absolute value of the critical current also depends strongly on the magnetization orientation. It only depends on the in-plane magnetization component along the current direction. We demonstrate that such dependence, in a suitably chosen voltage-driven regime, can lead to supercurrent induced splitting of the magnetic easy axis of the ferromagnet. We show that this effect may lead to stabilization of a four-fold degenerate ferromagnetic state, which is in sharp contrast to the conventional two-fold degenerate easy-axis ferromagnetic state.

The paper is organized as follows. In Sec. II we derive a CPR for the S/F/S junction atop a topological insulator surface starting from the quasiclassical Green function formalism.  This is followed by a discussion of the magnetization dynamics of such systems in Sec III. Next, in Sec IV, we discuss the stabilization of the four-fold degenerate ferromagnetic state. Finally, we conclude in Sec. V. 

\section{Current-phase relation in a ballistic S/F/S junction on a 3D TI}

\begin{figure}[!tbh]
   \centerline{\includegraphics[clip=true,width=2.8in]{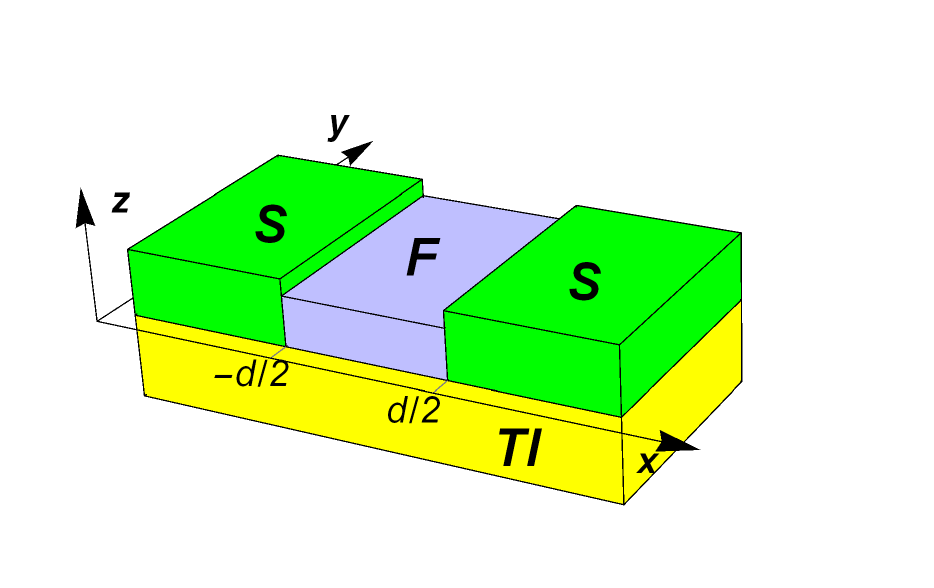}}
        \caption{Sketch of the system under consideration. Superconducting leads and a ferromagnetic interlayer are deposited on top of the TI insulator.}
 \label{sketch}
 \end{figure}

The sketch of the system under consideration is presented in Fig.~\ref{sketch}. Two conventional s-wave superconductors and a ferromagnet are deposited on top of a 3D TI insulator to form a Josephson junction. 

First of all, we consider a current-phase relation of a Josephson junction. The interlayer of the junction consists of the TI conducting surface states with a ferromagnetic layer on top of it. It is assumed that the magnetization $\bm M(\bm r)$ of the ferromagnet induces an effective exchange field $\bm h_{eff}(\bm r) \sim \bm M (\bm r)$ in the underlying conductive surface layer. The Hamiltonian that describes the TI surface states in the
presence of an in-plane exchange field $\bm h_{eff}(\bm r)$ reads:
\begin{equation}
\hat H=\int d^2 \bm r' \hat \Psi^\dagger (\bm r')\hat H(\bm r')\hat \Psi(\bm r')
\label{H},
\end{equation}
\begin{equation}
\hat H(\bm r)=-iv_F (\bm \nabla \times \bm e_z)\hat {\bm \sigma}+\bm h_{eff}(\bm r)\hat {\bm \sigma} -\mu
\label{h},
\end{equation}
where $\hat \Psi=(\Psi_\uparrow, \Psi_\downarrow)^T$, $v_F$ is
the Fermi velocity, $\bm e_z$ is a unit vector normal to the surface
of TI, $\mu$ is the chemical potential, and $\hat {\bm \sigma}=(\sigma_x, \sigma_y, \sigma_z)$ is a vector of
Pauli matrices in the spin space. It was shown \cite{Zyuzin2016,Bobkova2016} that in the quasiclassical approximation $(h_{eff},\varepsilon,\Delta) \ll \mu$ the Green's function has the following spin structure: $\check g (\bm n_F, \bm r, \varepsilon) = \hat g (\bm n_F, \bm r, \varepsilon) (1+\bm n_\perp \bm \sigma)/2$, where $\bm n_\perp = (n_{F,y},-n_{F,x},0)$ is the unit vector perpendicular to the direction of the quasiparticle trajectory $\bm n_{F} = \bm p_F/p_F$ and $\hat g$ is the {\it spinless} $4 \times 4$ matrix in the particle-hole and Keldysh spaces containing normal and anomalous quasiclassical Green's functions. The spin structure above reflects the fact that the spin and momentum of a quasiparticle at the surface of the 3D TI are strictly locked and make a right angle. Following standard procedures\cite{Eilenberger1968,Usadel1970} it was demonstrated\cite{Zyuzin2016,Bobkova2016,Hugdal2017} that the {\it spinless} retarded Green's function $\hat g(\bm n_F, \bm r, \varepsilon)$ obeys the following transport equations in the ballistic limit:
\begin{eqnarray}
-i v_F \bm n_F \hat \nabla \hat g = \Bigl[ \varepsilon \tau_z - \hat \Delta, \hat g \Bigr]_\otimes,
\label{eilenberger}
\end{eqnarray}
where $[A,B]_\otimes = A\otimes B -B \otimes A$ and $A \otimes B = \exp[(i/2)(\partial_{\varepsilon_1} \partial_{t_2} -\partial_{\varepsilon_2} \partial_{t_1} )]A(\varepsilon_1,t_1)B(\varepsilon_2,t_2)|_{\varepsilon_1=\varepsilon_2=\varepsilon;t_1=t_2=t}$. $\tau_{x,y,z}$ are Pauli matrices in particle-hole space with $\tau_\pm = (\tau_x \pm i \tau_y)/2$. $\hat \Delta = \Delta(x)\tau_+ - \Delta^*(x)\tau_-$ is the matrix structure of the superconducting order parameter $\Delta(x)$ in the particle-hole space. We assume $\Delta(x)=\Delta e^{-i\chi/2}\Theta(-x-d/2)+\Delta e^{i\chi/2}\Theta(x-d/2)$. The spin-momentum locking allows for including $\bm h_{eff}$ into the gauge-covariant gradient $\hat \nabla \hat A = \bm \nabla \hat A + (i/v_F)[(h_x \bm e_y - h_y \bm e_x)\tau_z, \hat A]_\otimes$.

Eq.~(\ref{eilenberger}) should be supplemented by the normalization condition $\hat g\otimes \hat g = 1$ and the boundary conditions at $x=\mp d/2$. As we assume that the Josephson junction is formed at the surface of the TI, the superconducting order parameter $\Delta$ and $\bm h_{eff}$ are effective quantities induced in the surface states of TI by proximity to the superconductors and a ferromagnet. In this case there are no reasons to assume existence of potential barriers at the $x=\mp d/2$  interfaces and we consider these interfaces as fully transparent.  In this case the boundary conditions are extremely simple and are reduced to continuity of $\hat g$ for a given quasiparticle trajectory at the interfaces.

To obtain the simplest sinusoidal form of the current-phase relation we linearize Eq.~(\ref{eilenberger}) with respect to the anomalous Green's function. In this case the retarded component of the Green's function $\hat g^R = \tau_z + f^R \tau_+ + \tilde f^R \tau_-$. The anomalous Green's function obeys the following equation:
\begin{eqnarray}
-\frac{1}{2}i v_F \partial_x f^{R} +  \bm h_{eff} \bm n_\perp f^{R} =  \varepsilon f^{R} - \Delta (x).
\label{eq:f}
\end{eqnarray}
Equation for $\tilde f^R$ is obtained from Eq.~(\ref{eq:f}) by $\bm v_F \to -\bm v_F$, $\Delta \to -\Delta$ and $\chi \to -\chi$.

The solution of Eq.~(\ref{eq:f}) satisfying asymptotic conditions $f^{R} \to (\Delta/\varepsilon)e^{\pm i \chi/2}$ at $x \to \pm \infty$ and continuity conditions at $x=\mp d/2$ takes the form [the solution is written for $x \in (-d/2,d/2)$, the solution in the superconducting leads is also found, but it is not required for finding the Josephson current]:
\begin{eqnarray}
f^R_{\pm} = \frac{\Delta e^{\mp i \chi/2}}{\varepsilon}\exp\Bigl[{\frac{\mp 2i (\bm h_{eff}\bm n_\perp - \varepsilon)(d/2 \pm x)}{v_x}}\Bigr],\nonumber \\
\tilde f^R_{\pm} = -\frac{\Delta e^{\mp i \chi/2}}{\varepsilon}\exp\Bigl[{\frac{\mp 2i (\bm h_{eff}\bm n_\perp - \varepsilon)(d/2 \mp x)}{v_x}}\Bigr],
\label{f_sol}
\end{eqnarray}
where the subscript $\pm$ corresponds to the trajectories ${\rm sgn}\! ~v_x =\pm 1$.  

The density of electric current along the $x$-axis is
\begin{eqnarray}
j_x = -\frac{e N_F v_F}{4} \int \limits_{-\infty}^{\infty} d \varepsilon \int \limits_{-\pi/2}^{\pi/2} \frac{d \phi}{2 \pi} \cos \phi \times \nonumber \\
\Bigl[(g^R_+ \otimes \varphi_+ - \varphi_+ \otimes g^A_+)-(g^R_- \otimes \varphi_- - \varphi_- \otimes g^A_-)\Bigr],
\label{current}
\end{eqnarray}
where $\phi$ is the angle, which the quasiparticle trajectory makes with the $x$-axis. $\varphi_{\pm}$ is the distribution function corresponding to the trajectories ${\rm sgn}\!~v_x =\pm 1$. 

Here we consider the voltage-biased junction. In principle, in this case the electric current through the junction consists of two parts: the Josephson current $j_s$ and the normal current $j_n$. The Josephson current is connected to presence of the nonzero anomalous Green's functions in the interlayer and takes place even in equilibrium. Here we assume the low applied voltage regime $eV/(k_B T_c) \ll 1$. In this case the deviation of the distribution function from equilibrium is weak and can be disregarded in calculation of the Josephson current: $\varphi_+ = \varphi_- = \tanh (\varepsilon/2T)$. Exploiting the normalization condition one can obtain $g^{R}_{\pm} \approx 1 - f^{R}_{\pm} \tilde f ^{R}_\pm/2$. Taking into account that $g^A_{\pm} = -g^{R*}_{\pm}$ we find the following final expression for the Josephson current:
\begin{eqnarray}
j_s = j_c \sin (\chi - \chi_0), \label{Josephson_CPR}\\
j_c = ev_F N_F T \sum \limits_{\varepsilon_n >0} \int \limits_{-\pi/2}^{\pi/2} d \phi \cos \phi \frac{\Delta^2}{\varepsilon_n^2}  \times \nonumber \\
\exp[-\frac{2\varepsilon_n d}{v_F \cos \phi}] \cos [\frac{2h_xd \tan \phi}{v_F}], \label{critical_current} \\
\chi_0 = 2 h_y d/v_F \label{chi_0},
\label{josephson_final}
\end{eqnarray}
where $\varepsilon_n = \pi T(2n+1)$. At high temperatures $T \approx T_c \gg \Delta$ the main contribution to the current comes from the lowest Matsubara frequency and Eq.~(\ref{critical_current}) can be simplified further 
\begin{eqnarray}
j_c = j_b \int \limits_{-\pi/2}^{\pi/2} d \phi \cos \phi  \times \nonumber \\
\exp[-\frac{2\pi T d}{v_F \cos \phi}] \cos [\frac{2h_xd \tan \phi}{v_F}], 
\label{critical_current_T} 
\end{eqnarray}
where $j_b = ev_F N_F \Delta^2/(\pi^2 T)$. Similar expression has already been obtained for Dirac materials \cite{Hugdal2017}. The normal current is due to deviation of the distribution function from the equilibrium. However, for the system under consideration, where we assume the ferromagnet to be metallic, practically all the normal current flows through the ferromagnet because in real experimental setups the TI resistance should be much larger as compared to the resistance of the ferromagnet. As for the Josephson current, it is carried by Cooper pairs and is strongly suppressed inside the ferromagnetic layer because the exchange field there is typically much larger as compared to the induced exchange field $\bm h_{eff}$ in the TI surface layer. Therefore, it flows through the TI surface states and we can assume that it is equal to the total electric current flowing via the {\it TI surface states}.  

\section{Magnetization dynamics induced by a coupling to Josephson junction}

The dynamics of the ferromagnet magnetization can be described in the framework of the Landau-Lifshitz-Gilbert (LLG) equation
\begin{eqnarray}
\frac{\partial\bm M}{\partial t} = -\gamma \bm M \times H_{eff} + \frac{\alpha}{M_s} \bm M \times \frac{\partial\bm M}{\partial t},
\label{LLG}
\end{eqnarray}
where $M_s$ is the saturation magnetization, $\gamma$ is the gyromagnetic ratio and $H_{eff}$ is the local effective field. The electric current flowing through the TI surface states causes spin-orbital torque\cite{Yokoyama2010,Yokoyama2011,Mahfouzi2012,Chen2014} due to the presence of a strong coupling between a quasiparticle spin and momentum. In principle, if the ferromagnetism and spin-orbit coupling spatially coexist, this torque is determined by the total electric current flowing through the system. However, for the case under consideration only the supercurrent flows via the TI surface states, where the spin-momentum locking takes place. Therefore, only this supercurrent generates a torque acting on the magnetization. The normal current flows through the homogeneous ferromagnet, where we assume no spin-orbit coupling. Consequently, it does not contribute to the torque.

The torque caused by the supercurrent can be accounted for as an additional contribution to the effective field. In order to find this contribution we can consider the energy of the junction as a sum of the magnetic and the Josephson energies:
\begin{eqnarray}
E_{tot}=E_M + E_J,
\label{energy}
\end{eqnarray}
where $E_J = E_c [1-\cos(\chi-\chi_0)]$ with $E_c = \Phi_0 I_c/2 \pi$, $I_c = j_c S$ ($S$ is the junction area) and $\chi = 2eVt$ in the presence of the applied voltage. is the Josephson energy. $E_M = -K V_F M_y^2/2M_s^2$ is the uniaxial anisotropy energy with the easy axis assumed to be along the $y$-axis. $V_F$ is the volume of the ferromagnet. The effective field $H_{eff} = -(1/V_F)(\delta E_{tot}/\delta \bm M)$ and takes the form:
\begin{eqnarray}
\frac{H_{eff,x}}{H_F} = \Gamma \bigg[\int_{-\pi/2}^{\pi/2} e^{-\tilde{d}/\cos\phi} \sin \phi  \sin \bigg( r m_{x}  \tan\phi\bigg) d\phi \bigg] \times \nonumber \\\bigg[1-\cos\bigg(\Omega_{J} t-r m_{y}\bigg)\bigg],~~~~~~~~~~\label{Hx} \\
\frac{H_{eff,y}}{H_F} = \Gamma \bigg[\int_{-\pi/2}^{\pi/2} e^{-\tilde{d}/\cos\phi} \cos \phi  \cos \bigg( r m_{x} \tan\phi\bigg) d\phi \bigg] \nonumber \\
\sin\bigg(\Omega_{J} t-r m_{y}\bigg)+m_{y},~~~~~~~~~~~~~\label{Hy} \\
H_{eff,z}	= 0,~~~~~~~~~~~~~~~~~~~~~~~~~~~~~~~~~~~~~~
\label{Hz}
\end{eqnarray}
where we have introduced the unit vector $\bm m = \bm M/M_s$, $\tilde d = 2 \pi T d/v_F$ is the dimensionless junction length, $\Gamma = \Phi_0 j_b S r /2 \pi K V_F$ is proportional to the ratio of the Josephson and magnetic energies,  $r=2dh_{eff}/v_F$, $\Omega_J = 2eV$ is the Josephson frequency and $H_F = \Omega_F/\gamma = K/M_s$.

The effective field consists of two contributions: the anisotropy field, which is directed along the easy axis, is represented by the last term in Eq.~(\ref{Hy}). The other terms are generated by the supercurrent. The same approach to study magnetization dynamics in voltage biased junctions has already been applied to systems with spin-orbit coupling in the interlayer \cite{Konschelle2009,Shukrinov2018}. The qualitative difference of our system based on the TI surface states from these works is that the critical current demonstrates strong dependence on the $x$-component of magnetization in our case, while it has been considered as independent on the magnetization direction earlier. This dependence leads to nonzero $H_{eff,x} \sim m_x$ at small $m_x$. This means that the easy $y$-axis can become unstable in a voltage-driven or current-driven junction, while this axis is always stable if the critical current does not depend on magnetization direction. Moreover, there is no difference for the system between $\pm m_x$-components of the magnetization. This leads to the remarkable fact that in a driven system the easy axis is not reoriented keeping two stable magnetization directions, as it has already been obtained before, but is split demonstrating  {\it four stable} magnetization directions. In the following section we study this effect in detail.

\section{Magnetization easy axis splitting}
 
It is obvious that $m_x=m_z=0$ is an equilibrium point of Eq.~(\ref{LLG}) with $H_{eff}$ determined by Eqs.~(\ref{Hx})-(\ref{Hz}). Now we investigate stability of this point. In the linear order with respect to $m_x$ the effective field can be written as follows:
\begin{eqnarray}
H_{eff,x} = A H_F m_x [1-\cos(\Omega_Jt-r )] , \nonumber \\
H_{eff,y} = H_F [1 + B \sin (\Omega_Jt-r )], 
\label{H_linear}
\end{eqnarray}
where $A>0$ and $B>0$ are constants.  By comparing Eqs.~(\ref{H_linear}) and (\ref{Hx}) it is seen that
\begin{eqnarray}
A=r \Gamma  \int \limits_{-\pi/2}^{\pi/2} e^{-\tilde d/\cos \phi}\sin \phi \tan \phi d \phi.
\label{A}
\end{eqnarray}

The LLG equation (\ref{LLG}) in the linear order with respect to $m_x$ and $m_z$ takes the form:
\begin{eqnarray}
\dot m_x = \frac{\gamma}{1+\alpha^2}\Bigl[ H_{eff,y}(m_z-\alpha m_x)+\alpha H_{eff,x} \Bigr], \nonumber \\
\dot m_z = \frac{\gamma}{1+\alpha^2}\Bigl[ -H_{eff,y}(m_x+\alpha m_z)+H_{eff,x} \Bigr], 
\label{LLG_linear}
\end{eqnarray}
while $\dot m_y = 0$. 

\begin{figure}[!tbh]
   \centerline{\includegraphics[clip=true,width=3.4in]{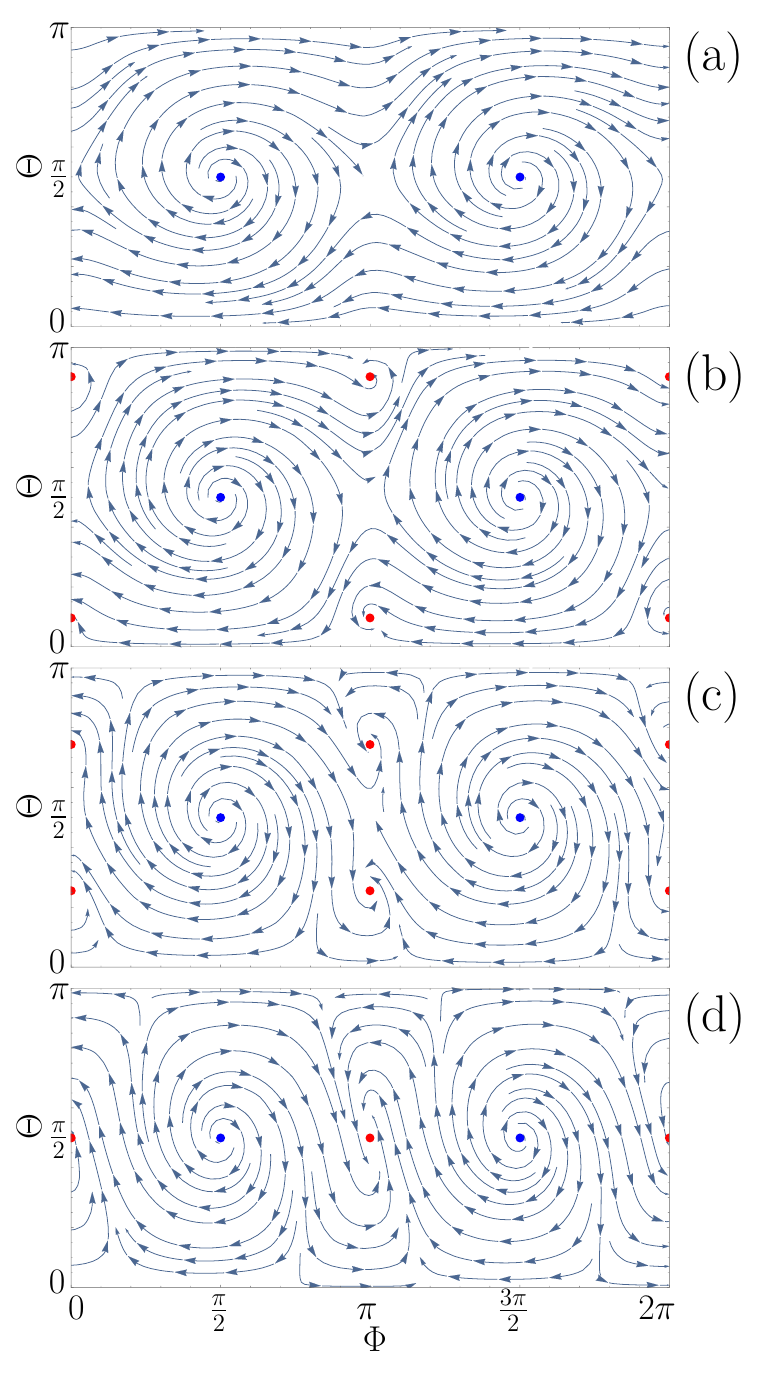}}
        \caption{Vector fields according to Eq.~(\ref{nonlinear}). (a) $A=0.90$ ($\Gamma = 1.26$); (b) $A=1.05$ ($\Gamma = 1.46$); (c) $A=1.25$ ($\Gamma = 1.84$); (d) $A=1.50$ ($\Gamma = 2.10$). $r=0.5$, $\tilde d = 0.3$, $\alpha = 0.25$ for all the panels. Blue points indicate unstable stationary solutions, and the stable solutions are marked by red points.}
 \label{vector}
 \end{figure}
 
\begin{figure}[!tbh]
   \centerline{\includegraphics[clip=true,width=3.4in]{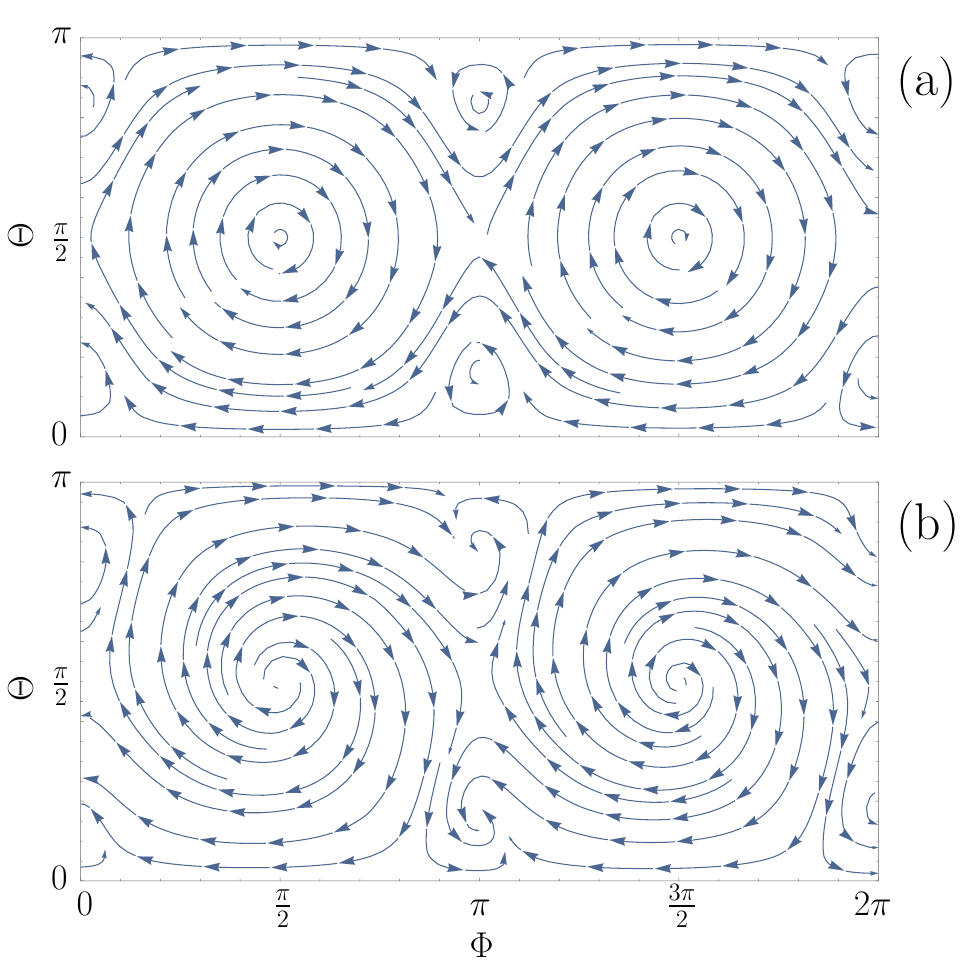}}
        \caption{(a) Vector field corresponding to the parameters of Fig.~\ref{evolution}, but for $\Omega_F/\Omega_J \to 0$. (b) The same as in panel (a), but for $\alpha = 0.25$ in order to demonstrate stability/instability of the stationary points.}
 \label{nonlinear_2}
 \end{figure}
 
\begin{figure}[!tbh]
   \centerline{\includegraphics[clip=true,width=3.7in]{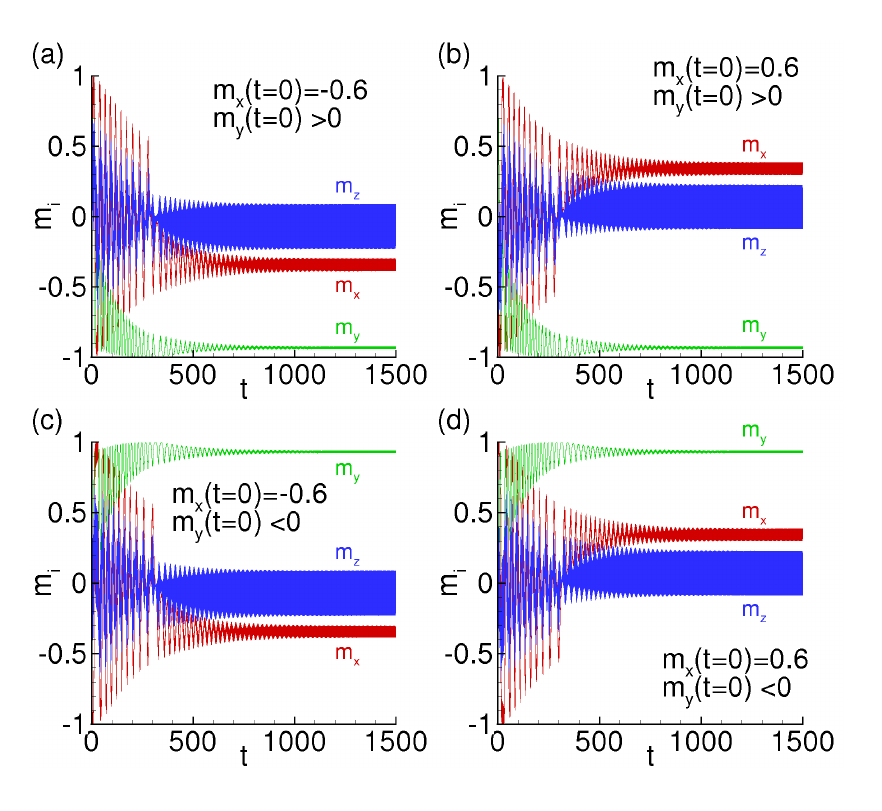}}
        \caption{Time evolution of the magnetization starting from different initial conditions. (a) $m_x(t = 0) = -0.6$, $m_y(t = 0) = 0.8$, (b) $m_x(t = 0) = 0.6$, $m_y(t = 0) = 0.8$, (c)
$m_x(t = 0) = -0.6$, $m_y(t = 0) = -0.8$ and (d) $m_x(t = 0) = 0.6$, $m_y(t = 0) = -0.8$. For all the panels we take $m_z(t = 0) = 0$. Four panels correspond to four possible stable states, which are reached by the system at large $t$. $\Gamma = 1.57$, $r=0.5$, $\tilde d = 0.3$, $\alpha = 0.01$, $\Omega_F/\Omega_J = 0.2$, time is measured in units of $\Omega_J^{-1}$.}
 \label{evolution}
 \end{figure}
 
One can estimate the parameter $\Omega_F/\Omega_J \sim \gamma H_F/(eI_cR_n)$ for 3D TI Josephson junctions. Taking $I_c R_n \sim 10^{-3}V$, as has been reported for $Nb/Bi_2Te_3/Nb$ Josephson junctions \cite{ Veldhorst2012}, and the easy-axis anisotropy field $H_F \sim 500 Oe$, what was reported for Py \cite{Beach2005,Beach2006}, we obtain $\Omega_F/\Omega_J \sim 3 \times 10^{-3}$. In the regime $\Omega_F/\Omega_J \ll 1$ the magnetization varies slowly at $t \sim \Omega_J^{-1}$, therefore we can average Eqs.~(\ref{LLG_linear}) over a Josephson period thus obtaining the following system:
\begin{eqnarray}
\dot m_x = \frac{\Omega_F}{1+\alpha^2}\Bigl[ m_z-\alpha (1-A)m_x \Bigr], \nonumber \\
\dot m_z = \frac{\Omega_F}{1+\alpha^2}\Bigl[ -(1-A)m_x-\alpha m_z \Bigr]. 
\label{LLG_linear_average}
\end{eqnarray}
The general solution of this system takes the form $m_{x(z)} = \sum \limits_{k=1,2}C_{k,x(z)}\exp[\lambda_kt]$. The equilibrium point $m_x=m_z=0$ becomes unstable under the condition ${\rm Re}\lambda_1>0$ or  ${\rm Re}\lambda_2>0$. One can obtain that it is realized at $A>1$.

It is rather difficult to make accurate estimates of the numerical value of $A$ for realistic parameters. The main problem is the absence of experimental data on the value of $h_{eff}$. However, if we take $K=500 J/m^3$ from the measurements \cite{Rusanov2004} on permalloy with very weak
anisotropy, $I_c = 10 \mu A$, $v_F \sim 10^5 m/c$ from \cite{Veldhorst2012} and the permalloy volume $d \times l \times w \sim 100 nm \times 10 nm \times 50 nm$, then we can obtain $A \sim r \Gamma \sim I_c h_{eff} /(v_F e K l w) \sim 0.4 - 8$ for $h_{eff} \sim 10 - 200 K$. Therefore, we can conclude that the range of $A$ values discussed in our  work should be experimentally accessible.

\begin{figure}[!tbh]
   \centerline{\includegraphics[clip=true,width=3.7in]{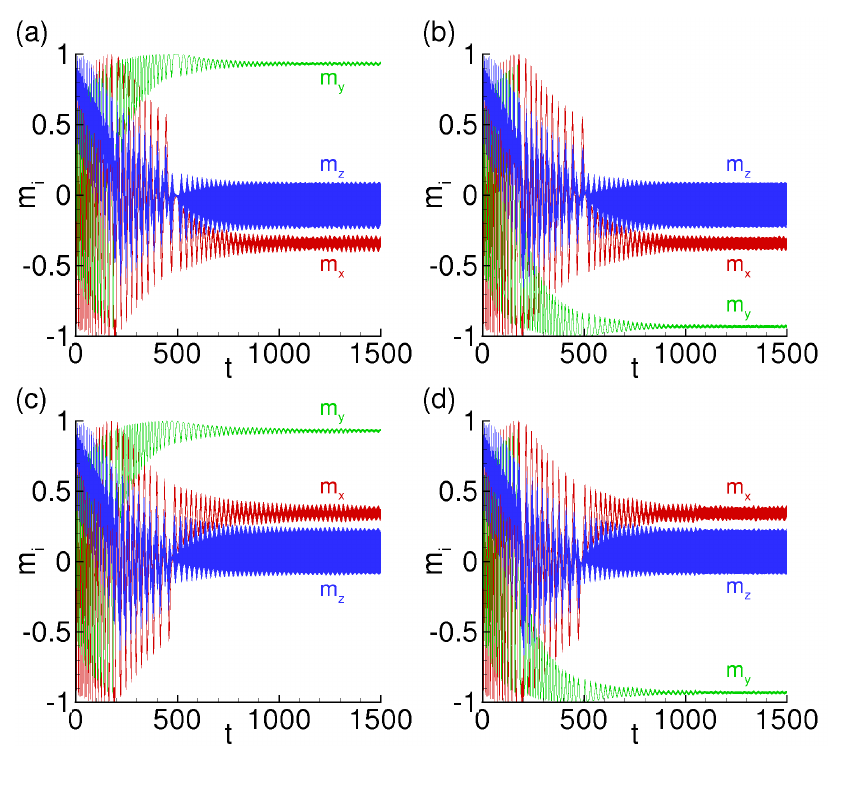}}
        \caption{Time evolution of the magnetization starting from unstable point with the initial condition $m_x=m_y=0$ and $m_{z}=1$ in the presence of noise. Four panels correspond to four possible stable states, which are reached by the system at large $t$. $\Gamma = 1.57$, $r = 0.5$, $\tilde{d}= 0.3$, $\alpha = 0.01$, $\Omega_{F}/\Omega_{J} = 0.2$, time is measured in units of $\Omega_J^{-1}$.}
 \label{noise}
 \end{figure}

\begin{figure}[!tbh]
   \centerline{\includegraphics[clip=true,width=3.0in]{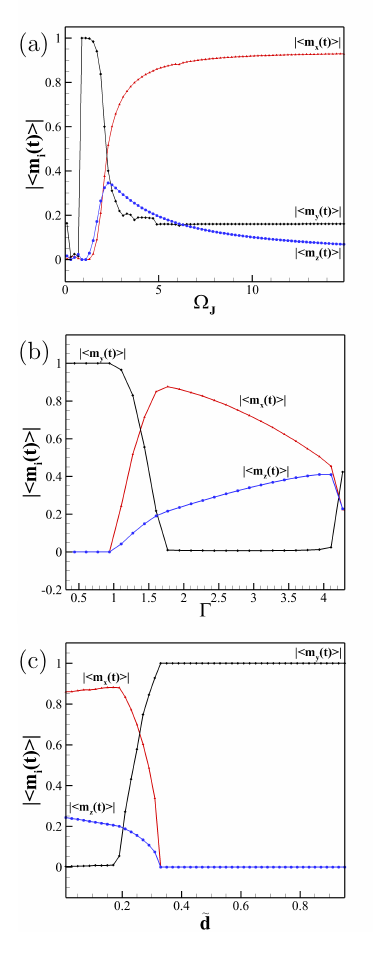}}
        \caption{(a) Averaged values of magnetization components at large times as functions of $\Omega_J/\Omega_F$. $\tilde d = 0.2$, $\Gamma = 1.62$. (b) The same as functions of $\Gamma$. $\tilde d = 0.2$, $\Omega_J/\Omega_F = 5$. (c) The same as functions of $\tilde d$. $\Omega_{F}/\Omega_{J} = 0.2$,  $\Gamma = 1.62$. For all the panels $r = 0.5$, $\alpha = 0.01$.}
 \label{parameters}
 \end{figure}

Now we turn to study the stationary points of the magnetization and their stability. Beyond the linear approximation (with respect to $m_x$ and $m_z$)  it is convenient to parametrize the magnetization as $\bm m = (\sin \Theta \cos \Phi, \cos \Theta, \sin \Theta \sin \Phi)$. Then from LLG equation one obtains:
\begin{eqnarray}
\dot \Theta = \frac{\gamma}{1+\alpha^2}\Bigl[ -\alpha H_{eff,y} \sin \Theta + \nonumber \\
H_{eff,x}(\sin \Phi + \alpha \cos \Theta \cos \Phi) \Bigr], \nonumber \\
\dot \Phi \sin \Theta = \frac{\gamma}{1+\alpha^2}\Bigl[ -H_{eff,y} \sin \Theta + \nonumber \\
H_{eff,x}(-\alpha \sin \Phi + \cos \Theta \cos \Phi)\Bigr].
\label{nonlinear}
\end{eqnarray}
At $\Omega_F/\Omega_J \to 0$ effective fields $H_{eff,x,y}$ determined by Eqs.~(\ref{Hx}), (\ref{Hy}) should be averaged over a Josephson period $H_{eff,x,y} \to \langle H_{eff,x,y} \rangle$. The stationary points are to be found as solutions of Eqs.~(\ref{nonlinear}) corresponding to $\dot \Theta = \dot \Phi = 0$.

Fig.~\ref{vector} shows vector fields in the plane $0 \leq \Phi < 2 \pi$, $0 \leq \Theta < \pi$ according to Eq.~(\ref{nonlinear}) 
at four different values of $A$. The stationary solutions are indicated by color points. The blue points correspond to unstable stationary solutions, while the red points indicate the stable magnetization directions. The Gilbert damping constant $\alpha = 0.25$. We have chosen such a large unrealistic value of the Gilbert constant in order to clearly show the stability/instability of the stationary points because for $\alpha = 0.01$, which is more appropriate for a realistic situation, stability/instability of a solution is not clearly seen in the figure [compare Figs.~\ref{nonlinear_2}(a) and (b)], although in fact the topology of the vector field is not changed. Fig.~\ref{vector}(a) represents the regime $A<1$, when the only stable solutions $\bm m^{st}$ are $m_y^{st}=\pm 1$, what corresponds to upper and bottom horizontal lines in the figure. Panels (b) and (c) demonstrate the vector fields in the regime of not very large $A>1$. Four stable red points are clearly seen. Upon further increase of $A$ the stable points get closer to $\Theta = \pi/2$ and finally merge into two stable points at some $A_{crit}$, as it is shown in Fig.~\ref{vector}(d). Therefore, there exists a finite range of $1<A<A_{crit}$, where the ferromagnet has four stable magnetization directions in the voltage-biased regime considered here. From Fig.2 it is seen that all the stationary points correspond to $m_z = \pm 1$ or $m_z=0$. The stationary points $m_z=\pm 1$ are always unstable. Let us consider the stationary points  corresponding to $m_z=0$, that is $\Phi=0,\pi$. It is obvious that in order to have four stable points $|m_x^{st}|$ and $|m_y^{st}|$ should be nonzero simultaneously. Substituting $m_z=0$ into Eq.~(\ref{nonlinear}) and taking into account that $\langle H_{eff,y} \rangle=H_F m_y$, we obtain that $m_x^{st}$ can be determined from the simple nonlinear equation $m_x = \langle H_{eff,x} \rangle/H_F$. This equation always has the solution $m_x = 0$, but at $1<A<A_{crit}$ it also has the second nonzero solution $m_x^{st}$. In this situation there are four stable points $\bm m^{st}=(\pm |m_x^{st}|,\pm |m_y^{st}|,0)$.

Further in Fig.~\ref{evolution} we demonstrate the full time evolution of the magnetization $\bm m$ obtained from the numerical solution of the LLG equation. It is seen that starting from different initial conditions it is possible to reach all four stable magnetization solutions. The results are obtained at $\Omega_F/\Omega_J = 0.2$, but the averaged values of magnetization at large times are in good agreement with the results for stable points obtained in the limit $\Omega_F/\Omega_J \ll 1$, which are demonstrated in Fig.~\ref{nonlinear_2}(a) for the same parameters $\Gamma$, $r$, $\alpha$ and $d$. Fig.~\ref{nonlinear_2}(b) only differs from (a) by the value of $\alpha = 0.25$. While the topology of the vector fields presented in panels (a) and (b) is the same, the stability/instability of all the stationary points is more clearly seen for larger values of the damping constant $\alpha$. At finite values of $\Omega_F/\Omega_J$ the magnetization oscillates around the vector trajectory presented in Fig.~\ref{nonlinear_2} and the amplitude of the oscillations is suppressed at $\Omega_F/\Omega_J \to 0$.

In order to show that the system under consideration can demonstrate spontaneous behavior we investigate the system evolution starting from one of unstable points $m_z = \pm 1$. A small noise is introduced to the system in order to allow for leaving the unstable equilibrium point. From the vector fields represented in Fig.~\ref{nonlinear_2}(a) it is seen that at small values of $\alpha$ the system finally comes to one of  the four stable states with practically equal probabilities. It is shown in Fig.~\ref{noise}, where different panels correspond to all the possible final states.

Fig.~\ref{parameters} demonstrates the behavior of the absolute values of averaged magnetization at $t \to \infty$ depending on essential parameters of the system. The dependence on $\Omega_J/\Omega_F$ is represented in Fig.~\ref{parameters}(a). It is seen that at $\Omega_J/\Omega_F \gg 1$ $|\langle m_i \rangle |$ tend to constant values and, in particular, $|\langle m_z \rangle | \to 0$, as it follows from our analysis of stationary points of Eqs.~(\ref{nonlinear}).

The dependence on $\Gamma$ is plotted in Fig.~\ref{parameters}(b). $\Gamma$ is linearly proportional to $A$. For this reason one can explicitly see in this panel the range of $A$ where four stable limiting magnetization directions exist: it corresponds to the regions, where $|\langle m_x \rangle |$ and $|\langle m_y \rangle |$ are nonzero simultaneously.

Panel (c) of Fig.~\ref{parameters} represents the dependence of $|\langle m_i \rangle |$ on the junction length. Analogously to the previous panel, the range of existence of four stable limiting magnetization directions is also clearly seen. The dependence on $r$ is qualitatively very similar to the dependence on $\Gamma$, therefore we do not represent it.  Figs.~\ref{parameters}(b)-(c) also provide the optimal range of parameters $\Gamma$ and $d$ for experimental realization of the easy axis splitting. The effect can be experimentally investigated, for example, by measuring the magnetic field pattern away from the nanomagnet.

\section{Conclusions}

In this work we study a S/F/S Josephson junction atop a topological insulator and discuss the possibility of electrical control of magnetization in such a device. Our analysis,
which combines microscopic Keldysh Green function tehnique for obtaining the Josephson current with phenomenological Landau-Lifshitz-Gilbert equations for studying magnetization dynamics, shows that the property of full spin momentum locking can
lead to destabilization of a transverse easy magnetization axis $m_y$ in such systems in the presence of appropriate voltage or
current bias. Such an instability,  in its turn, results in a ferromagnet with two easy axes allowing for {\it four stable} magnetization directions instead of usual two. Switching between these states by means of voltage or current impulses is of interest from the applied point of view.

\section*{Acknowledgements}
The work of I.V.B. and A.M.B. was carried out within the state task of ISSP RAS. The reported study was partially funded by the RFBR
research projects  18-02-00318 and 18-52-45011-IND.  Numerical calculations have been made in the framework of the RSF project 18-71-10095. K.S. thanks DST for support through INT/RUS/RFBR/P-314.

\end{document}